\documentclass[]{iau}

\usepackage{amsmath}
\usepackage{graphicx}
\usepackage{xcolor}
\usepackage{multirow}
\usepackage{csquotes}

\usepackage{natbib}

\usepackage{caption}
\usepackage{subcaption}

\begin{document}

\lefttitle{E. Meyer}
\righttitle{Extragalactic Jets from Radio to Gamma-rays}

\jnlPage{1}{7}
\jnlDoiYr{2023}
\doival{10.1017/xxxxx}
\aopheadtitle{Proceedings IAU Symposium}
\volno{375}
\editors{Y.\,Liodakis eds.}

\title{Extragalactic Jets from Radio to Gamma-rays}

\author{E.~Meyer$^1$, A.~Shaik$^1$, K. Reddy$^2$, M. Georganopoulos$^{1,3}$}
\affiliation{$^1$\,University of Maryland Baltimore County, 1000 Hilltop Cir, Baltimore MD 21250, USA -- $^2$\,School of Earth and Space Exploration, Arizona State University, Tempe, AZ 85287, USA -- $^3$NASA Goddard Space Flight Center, Greenbelt, MD USA}

\begin{abstract}
Despite the fact that jets from black holes were first understood to exist over 40 years ago, we are still in ignorance about many primary aspects of these systems -- including the radiation mechanism at high energies, the particle makeup of the jets, and how particles are accelerated, possibly to energies as high as 100 TeV and hundreds of kpc from the central engine. We focus in particular on the discovery (and mystery) of strong X-ray emission from radio jets on kpc-scales, enabled by the unequaled high resolution of the \emph{Chandra} X-ray observatory. We review the main evidence for and against the viable models to explain this X-ray emission over the last 20 years. Finally, we present results of a recent study on the X-ray variability of kpc-scale jets, where we find evidence that between 30-100\% of the X-ray jet population is variable at the tens-of-percent level.   The short ($\sim$years) variability timescale is incompatible with the IC/CMB model for the X-rays and implies extremely small structures embedded within the kpc-scale jet, and thus requires a reconsideration of many assumptions about jet structure and dynamics.  
\end{abstract}

\begin{keywords}
AGN, Blazar, Jets, Variability
\end{keywords}

\maketitle
\section{Introduction}
Active galactic nuclei (AGN) are powered by the supermassive black holes (SMBH) which reside in the centers of all relatively massive galaxies \citep{kormendy2013}. A small percentage of these sources eject bipolar, collimated jets of relativistic plasma which emit brightly at radio and sometimes optical and X-ray frequencies kiloparsecs (kpc) away from the central engine. Resolved radio jets are often separated into two classes based on radio morphology: Fanaroff \& Riley (1974) class I (FRI) jets have plume-like jets and are dominated by emission near the core, while FRII jets are highly collimated and are dominated by bright hotspots where the jet impacts into the intergalactic medium (IGM). Historically, this classification system was also associated with a difference in jet power, with FRI and FRII jets representing low- and high-power jets respectively. However, more recent studies have discovered low-power FRII galaxies \citep{mingo2019} and large population studies suggest a large range in jet power overlap \citep{keenan2021}.

One of the major discoveries by the \textit{Chandra} X-ray Observatory has been the detection of X-rays from radio jets on kpc-scales \citep{chartas2000,schwartz2000,harris2006,worrall2009,marshall2018}. Assuming a leptonic jet model, the X-ray emission of most low-power or FRI class sources, including nearby sources such as M87 \citep{harris2003_sed} and Cen A \citep{hardcastle2007}, are often (but not always) well-described by synchrotron radiation from a single electron population which extends from radio to optical and X-ray energies. However in many (typically the most powerful) quasar-hosted jets, the X-ray emission is too high, and the spectral index too hard, to be attributed to the low-energy synchrotron component. A classic example (and the first X-ray source \emph{Chandra} observed) of PKS~0637-752 is shown in Figure~1. The observation of `anomalously' bright and hard X-rays has since been extended to virtually all powerful FRII jets and even some low-power FRI jets such as in the case of M84 \citep{meyer2018}. 

\begin{figure}
    \centering
    \includegraphics[width=\textwidth]{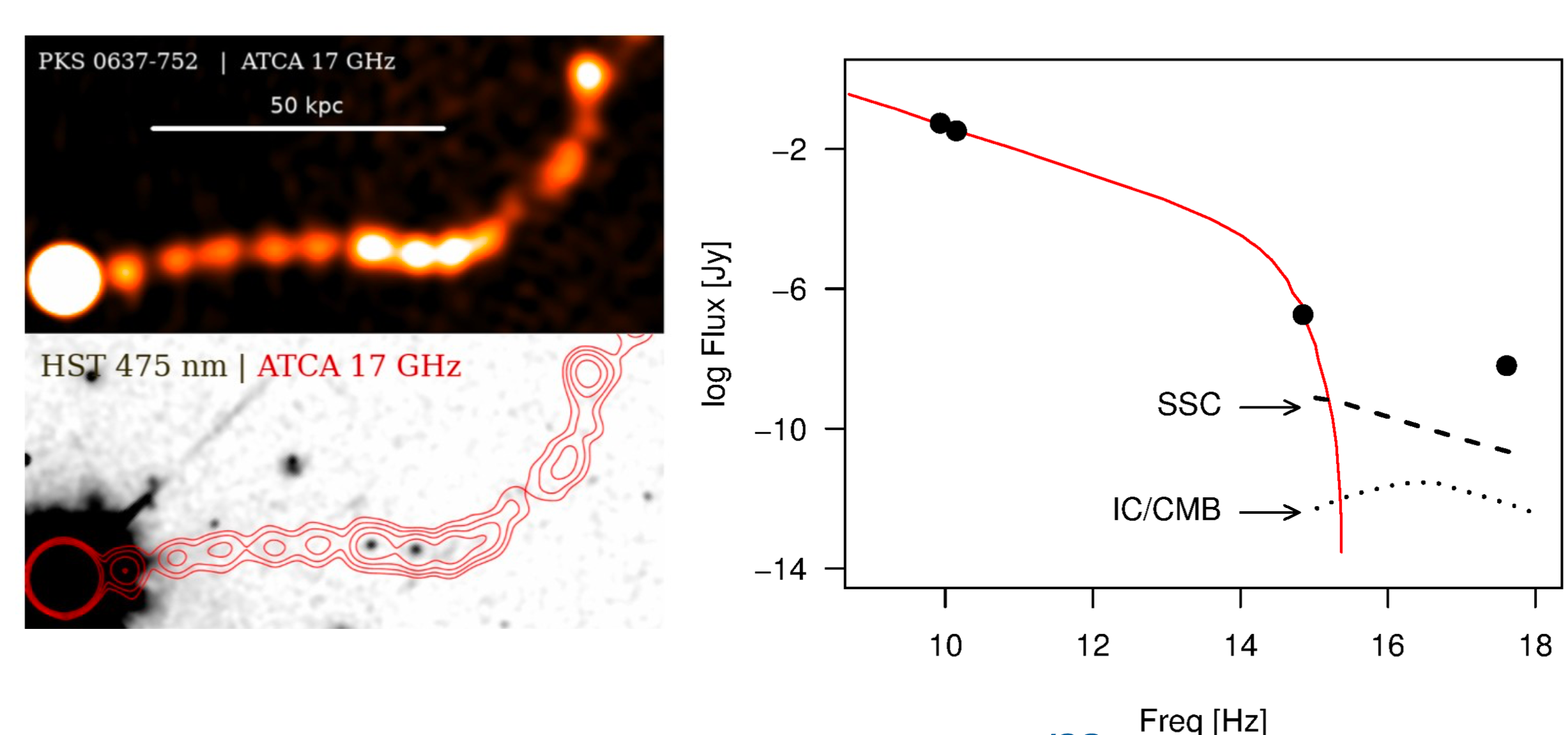}
    \caption{Upper left, an ATCA radio image of PKS~0637-752 (provided by Leith Godfrey), below, an HST image of the jet with inverted color scale and radio contours overlaid. Note the small knots of optical emission coinciding with the brightest radio knots. At right, an early SED of the bright knots of the jet, showing the very high level of X-ray emission and the difficulty of explaining it with either SSC or (unbeamed) IC/CMB. }
    \label{fig:pks0637}
\end{figure}

\section{The Rise and Fall of the IC/CMB model}
The most commonly adopted explanation for the `anomalous' X-ray emission from kpc-scale jets is inverse-Compton scattering of Cosmic Microwave Background (IC/CMB) photons by a still-relativistic jet. Under this model, high-energy electrons in the jet upscatter low-energy CMB photons, provided that the jet is both highly relativistic with high bulk Lorentz factors ($\Gamma > 10$)  on kpc-scales and closely aligned to our line-of-sight \citep[$\theta < 5^\circ$;][]{tavecchio2000,celotti2001}. This produces the high Doppler boosting required to reproduce the observed bright and hard-spectrum X-ray flux. However, this emission mechanism also sometimes predicts kinetic powers in excess of the Eddington limit \citep{atoyan2004} and is in many cases at odds with other more recent observations, including significantly higher polarization of the ultraviolet/X-ray component than expected \citep{cara2013} and jet-counterjet flux ratios which are inconsistent with IC/CMB predictions \citep{kataoka2008, clautice2016, hardcastle2016} . 

\begin{figure}
    \centering
    \begin{subfigure}[t]{0.48\textwidth}
    \centering
     \includegraphics[width=\textwidth]{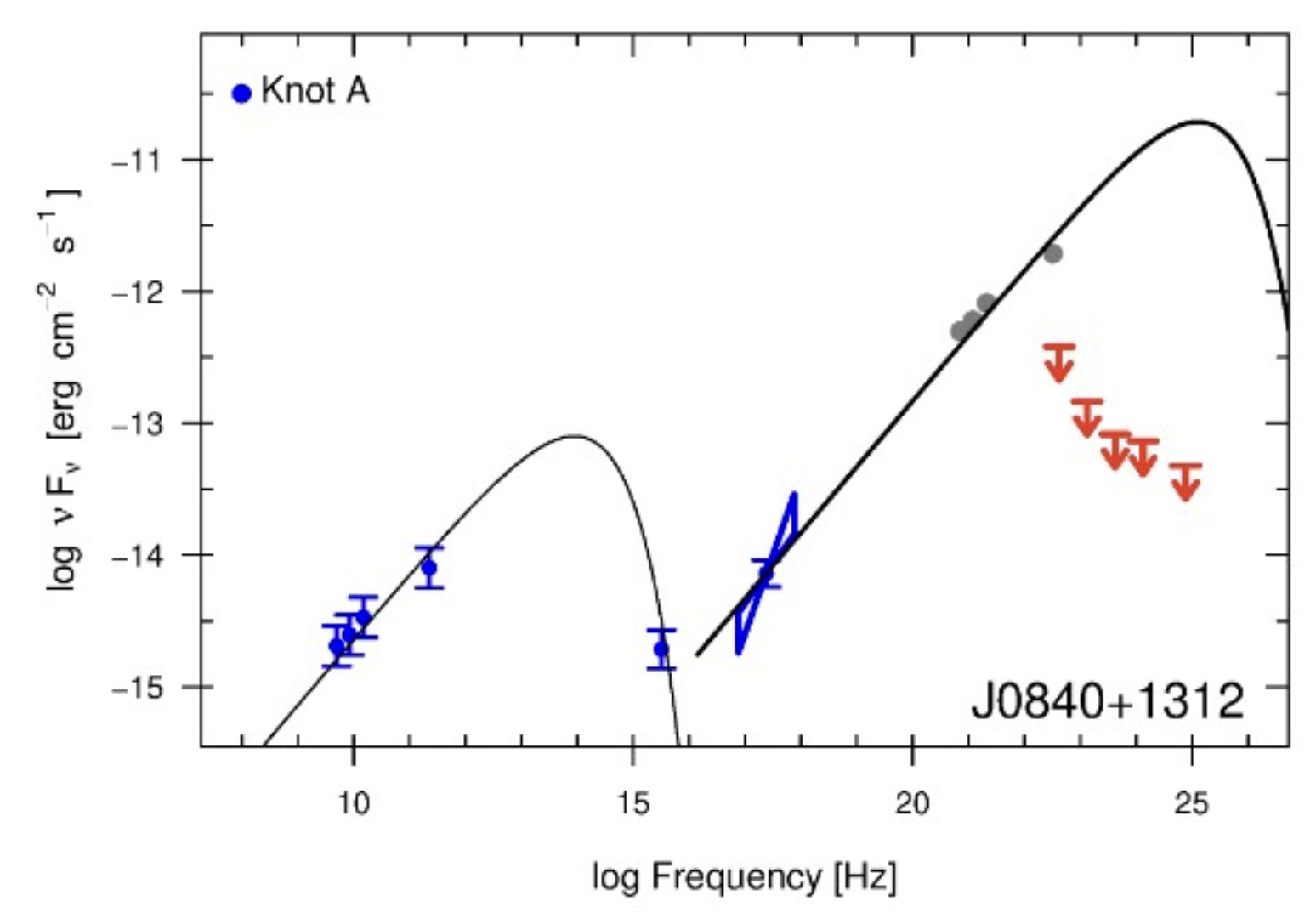}
    \end{subfigure}
    \hfill
   \begin{subfigure}[t]{0.48\textwidth}
   \centering
    \includegraphics[width=\textwidth]{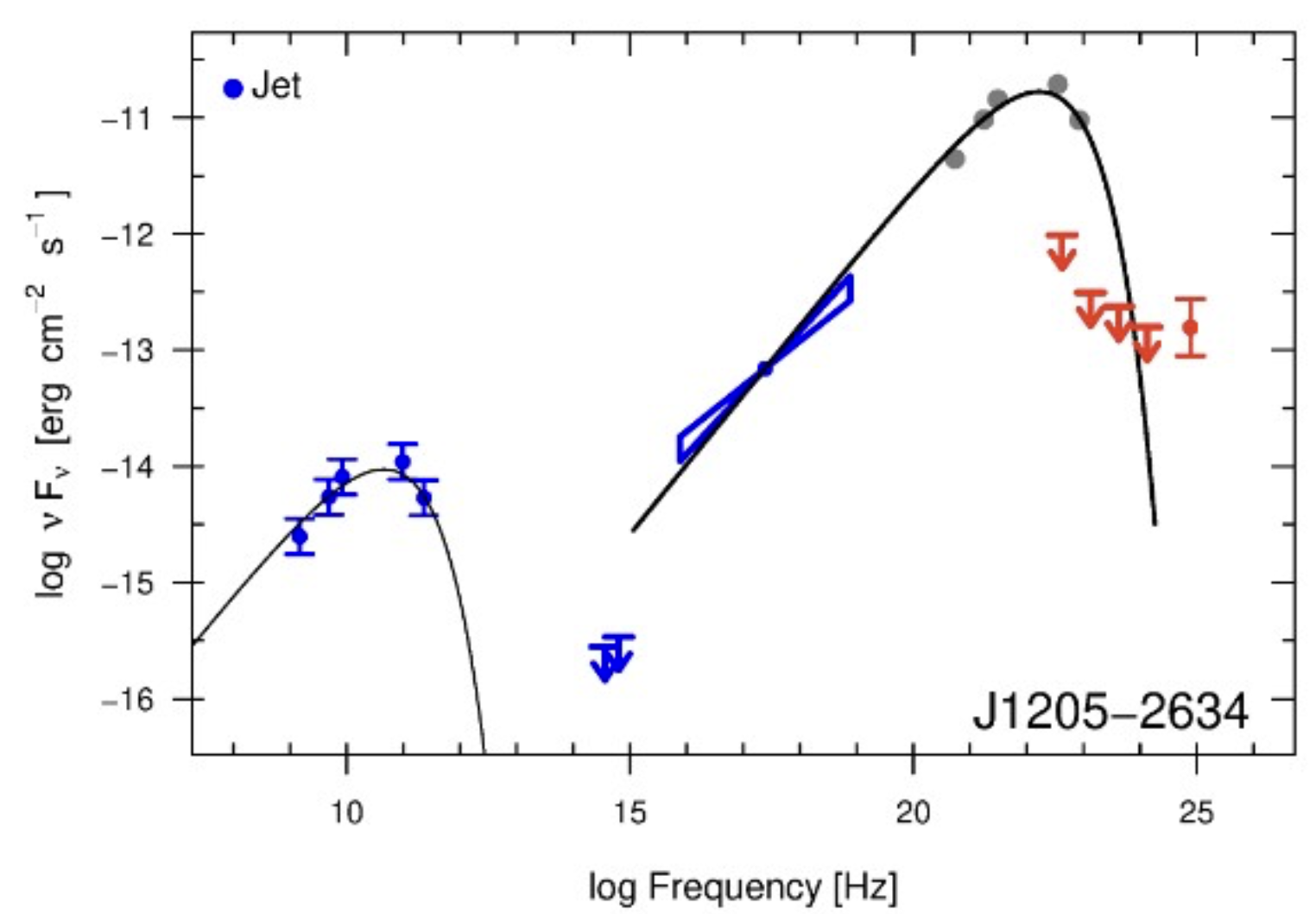}
    \end{subfigure}
    
    \caption{Example large-scale (kpc-scale) jet SEDs, with synchrotron (thin line) and IC/CMB (thick line) model curves. As shown, matching the X-ray emission unavoidably predicts a high level of GeV emission, which is in violation with the deep \emph{Fermi}/LAT upper limits (red). Taken from \cite{breiding2023}. }
    \label{fig:peter}
\end{figure}

Another major effort at testing the IC/CMB model has been through looking for the high levels of GeV emission that it predicts, first proposed before the launch of \emph{Fermi} by \cite{geo2006}. The analysis is complicated by the poor (fractions of a degree) resolution of \emph{Fermi} as it is impossible to spatially distingues the jet from the core, which is highly variable and is expected dominate over the jet especially at lower (MeV) energies. However, through the use of a recombined light-curve analysis, \cite{meyer2014} first showed that the IC/CMB model could be strongly ruled out through gamma-ray upper limits for the X-ray jet of 3C~273, before doing the same in PKS~0637-752 \citep{meyer2015,meyer2017}. Peter Breiding then widely applied this test to essentially all the `multiple spectral component' or MSC jets for his PhD thesis work \citep{breiding2017, breiding2023}, finding that 24/45 sources tested would require IC/CMB-predicted GeV emission significantly in violation of the gamma-ray limits (see example SEDs for two of the sources from \citet{breiding2023} in Figure~\ref{fig:peter}). 

Another difficulty for the IC/CMB model is the change in apparent morphology (knot location, size, etc) from radio to optical to X-ray. The simplest IC/CMB model requires cospatial emission in these different bands, since the radio-to-optical synchrotron spectrum is produced by the same electron energy distribution upscattering the CMB to X-ray energies (or if there is an offset, for the X-rays to persist beyond the radio since these arise from very low-energy electrons in the IC/CMB interpretation).  However, an extensive study of essentially all X-ray jets discovered to date finds significant offsets between X-rays and radio  on the order of  $\sim$1~kpc or more \citep[e.g.,][]{karthik+21,karthik+22}.

\section{Variability of X-ray Jets}
One consequence of the IC/CMB model is a low energy extension of the electron energy distribution. That is, in order for IC/CMB to be the dominant X-ray emission mechanism, there must be a significant population of particles at low-energies \citep[$\gamma_{min}\sim 10-20$;][]{celotti2001,geo2006}. Besides the considerable energy requirements of this extension, the implied radiative cooling times are extremely long (longer than the jet lifetimes), which means that under an IC/CMB X-ray mechanism we should absolutely not expect to see variability in X-ray emission on timescales within the lifetime of \textit{Chandra}.

It was thus surprising when \cite{marshall2010} examined the case of the nearby ($z = 0.035$) FRII Pictor A, a radio galaxy with a very long (arcminute-scale) radio jet and steep radio core, and a clearly visible counterjet. A knot in the jet of Pictor~A was seen to fade over a timescale of a few years with a reported significance of 3.4$\sigma$. \cite{hardcastle2016} and \cite{thimmappa2020} later confirmed these results at similar significances (\textit{p} $<$ 0.011). While variability had been observed by this point in nearby FRI jets \citep[most dramatically in the mid-2000s outburst of HST-1 in the jet of M87;][]{harris2003} variability on this timescale was completely unexpected. In the case of Pictor A, assuming minimum energy conditions and an emitting region the size of the jet cross-section, the synchrotron cooling timescale is on the order of 1200 years, implying that the X-rays arise from a region much smaller and with a magnetic field several times higher than equipartition.

\begin{figure}
    \centering
    \includegraphics[width=\textwidth]{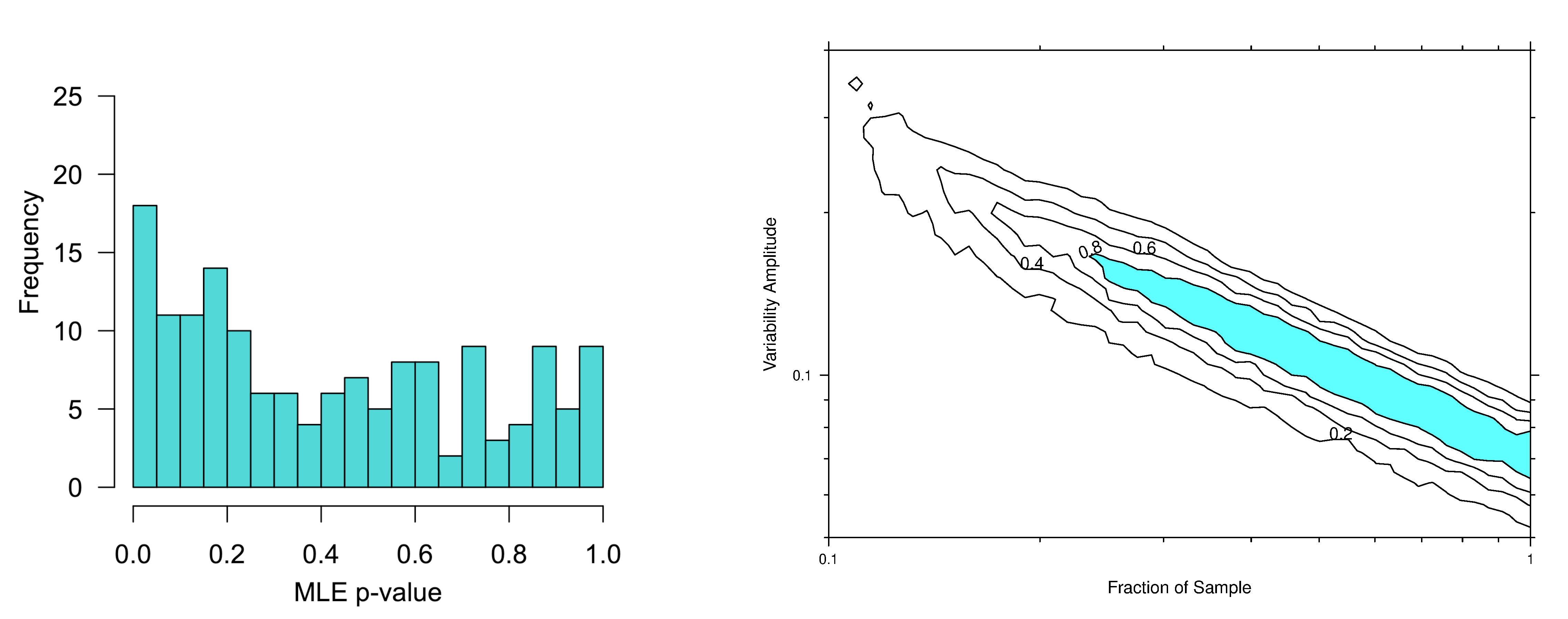}
    \caption{At left, the histogram of p-values from our likelihood model analysis of 155 jet regions from 53 total sources. Under the null hypothesis of steady emission, we expect a completely uniform (0,1) distribution of p-values. The global p-value associated with the excess of low values shown here is p=0.00019, suggesting variability in the jet population. At right, the results of a simulation to test the degeneracy between sample fraction of variability (assuming 1-f are steady emitters) and the typical variability amplitude (relative to mean). The shaded blue region shows the best agreement with the observational data, suggesting between 30-100\% of jets are variable, at a modest tens-of-percent scale. }
    \label{fig:var}
\end{figure}

\vspace{12pt}
\noindent
\subsection{New Results}
\vspace{5pt}

Over the past few years, we have conducted a comprehensive archival study of X-ray jets, to look for variability in other sources. Our sample of 53 comprises nearly all known X-ray jets imaged more than once by the \emph{Chandra} Advanced CCD Imaging Spectrometer (ACIS) instrument.  The average number of observations per source in our sample is 3.4, with a mean spacing of 2.6 years. Two sources (3C\,305 and 3C\,171) were excluded as the extended X-ray emission associated has been attributed to jet-driven gas \citep{hardcastle2012_3c305,hardcastle2010_3c171}; we also exclude the sources M87 and Centaurus A, which are already known to produce X-rays by synchrotron process \citep{feigelson1981,kraft2002,wilson2000}. Both M87 and Centaurus A have been far more deeply and frequently observed than typical of the remaining sample (with 50 and 43 distinct observations, respectively), and this is likely one of the reasons that X-ray variability has been reported in both cases.  With the exception of  Pictor A, variability has not been reported for any other source in our sample of 53.   Unlike the jets of M87 and Centaurus A, however, the X-ray emission in Pictor~A is clearly from a second emission component \citep{hardcastle2005}.

Our likelihood function is a straightforward application of Poisson statistics, with a null hypothesis of a steady source rate for each individual knot region, taking into account a varying background and various detector effects.  We computed for each knot a \emph{p}-value for the test of the the null hypothesis of a steady source rate, the distribution of which (for 155 test regions) is shown in Figure~\ref{fig:var}. Out of the full sample of 155 regions tested, 18 (12\%) have \emph{p}-values less than 0.05, suggesting significant variability in the intrinsic source rate.  The single-region \emph{p}-values for the are expected to follow a uniform $U(0,1)$ distribution under the null hypothesis of steady emission, so this is clearly about twice the value expected, and the excess of low p-values is apparent by eye in Figure~3.  This is also born out in the statistics: when we compare all 155 single-region \emph{p}-values to a $U(0,1)$ distribution using a one-sided Kolmogorov-Smirnov (KS) test, we obtain a global \emph{p}-value of 0.000196, indicating that the distribution is highly non-Uniform. This clearly indicates that the observations are not consistent with non-variable X-ray sources.

Figure~3 (right panel) shows the results of a simulation to try to constrain the typical scale of variability in our jet population, given a  degree of degeneracy between the typical variability amplitude and rate of variable sources existing in the population. It is not implausible that a subset of X-ray jets are steady X-ray emitters while the rest exhibit variability. For example, 30\% of the sample might be variable sources, with a characteristic scale of 50\% in amplitude, or 90\% of the sample might be variable with a lower (say, 10\%) characteristic amplitude of variability, and both of these scenarios might be consistent with the $p$-value distribution we observe. Our simulation compares the MLE $p$-value distribution of simulated populations varying in these two respects to our actual $p$-value distribution and reports the K-S test $p$-value for the comparison. Thus, the lighter-colored regions of the plot are most consistent with the data while black areas are not consistent with the data under a K-S test comparison.

\begin{figure}
    \centering
    \includegraphics[width=\textwidth]{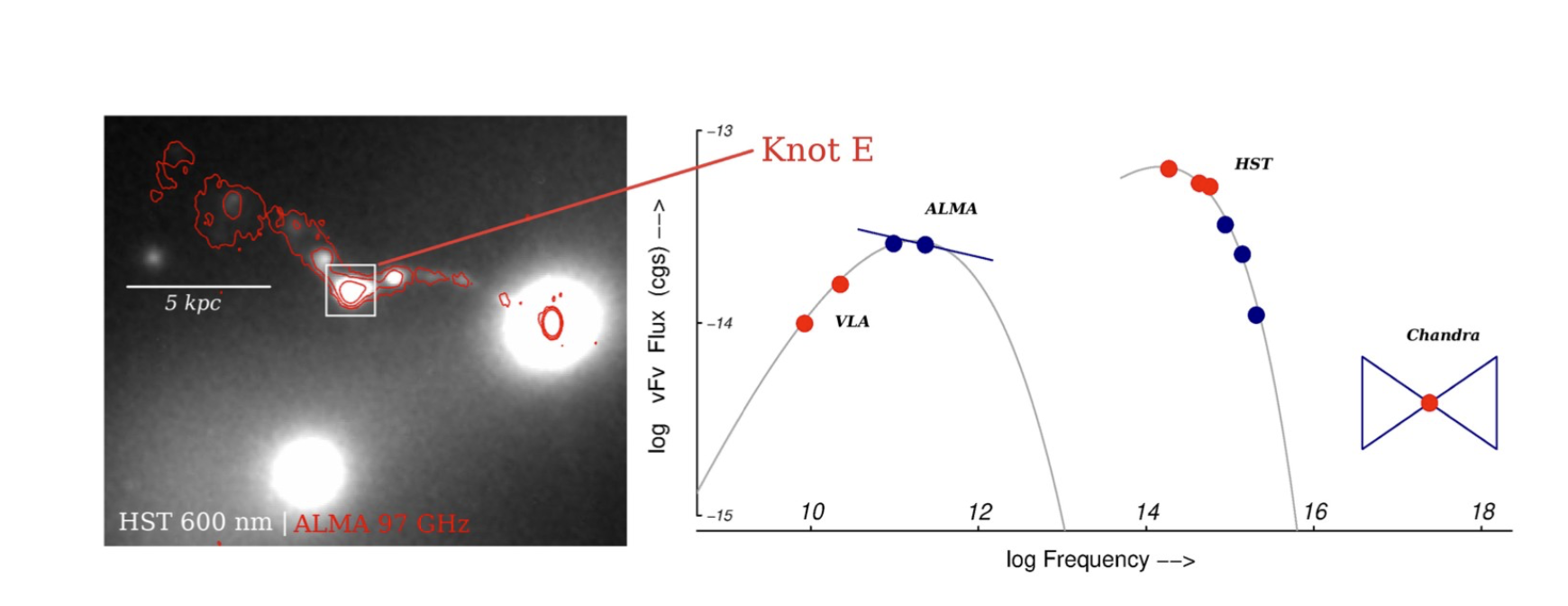}
    \caption{At left, an HST image of the jet of 3C 346, with ALMA band3 contours overlaid. At right, a detailed SED of knot E, showing at least 3 emission components in the jet, so far unexplained (Meyer et al., in prep.)}
    \label{fig:3c346}
\end{figure}

\section{Discussion}

As a result of this and the other observational lines of evidence,  the \emph{general} validity of IC/CMB model for large-scale jet X-ray emission must be called into question. On the other hand, IC/CMB does appear to be more consistent than alternatives in some cases, for example the few high-redshift jets with unusually faint or undetected radio emission \citep[e.g.][]{sim2016,mig2022}, and potentially in somewhat `unusual' cases at lower redshift. For example, we found that the 'plateau' or underlying steady-state gamma-ray emission in two X-ray jets (PKS 1510-089 and OJ 287) was compatible with the predicted IC/CMB level \citep{meyer2018}. However, these cases are likely outliers, with a particularly favorable alignment and unusually high jet speed at large distances.
 
The main alternative to the IC/CMB model is synchrotron radiation from a second electron energy distribution (EED), or abandoning the leptonic model in favor of a hadronic jet model \citep{petropoulou2017, meyer2018}. Synchrotron models allow us to avoid the 'uncomfortable' requirements of IC/CMB models (e.g., high bulk Lorentz factors, small viewing angle, and super-Eddington jet powers), but still have a largely ad-hoc appearance: multiple EEDs was not predicted by theory. Previously, the apparent co-spatiality of the radio and X-rays seemed consistent with a single population (i.e. IC/CMB) X-ray model, but recent research has shown that they are in fact \textit{not} co-spatial in most cases \citep{karthik+22}. Notably, these different mechanisms along with IC/CMB involve vastly different scenarios as far as jet matter content and energy distribution, jet power, and acceleration mechanisms, with important implications for a proper accounting of jet impacts on the environment.

The short timescales of X-ray variability observed in our study implies that the emitting regions are much smaller that the width of the jet (which is generally resolved in e.g., radio to be on kpc scales). By the light crossing time argument a flare event on the order of a year cannot occur in a region larger than a few parsecs; this requires highly localized particle acceleration and appears more consistent with magnetic reconnection than the usually assumed shock acceleration \citep{giannios2009}.  Interestingly, production of high energy photons from regions much smaller than the source physical size has been observed in very different celestial objects, from the Sun \citep{omodei2013} and the Crab Nebula \citep{abdo2011}, to blazars like PKS 1510-089, where variations over hours are seen in the VLBI radio core that has a light crossing time of the order of a year, a difference too large to explain through beaming \citep{marscher2010}. These observations are key to understanding the particle acceleration mechanism acting in these environments.

\bibliographystyle{iaulike}
\bibliography{main}

\begin{thebibliography}{}

\bibitem[{Abdo} et~al., 2011]{abdo2011}
{Abdo}, A.~A., {Ackermann}, M., {Ajello}, M., {Allafort}, A., {Baldini}, L.,
  {Ballet}, J., {Barbiellini}, G., {Bastieri}, D., {Bechtol}, K., {Bellazzini},
  R., {Berenji}, B., {Blandford}, R.~D., {Bloom}, E.~D., {Bonamente}, E.,
  {Borgland}, A.~W., {Bouvier}, A., {Brandt}, T.~J., {Bregeon}, J., {Brez}, A.,
  {Brigida}, M., {Bruel}, P., {Buehler}, R., {Buson}, S., {Caliandro}, G.~A.,
  {Cameron}, R.~A., {Cannon}, A., {Caraveo}, P.~A., {Casandjian}, J.~M.,
  {{\c{C}}elik}, {\"O}., {Charles}, E., {Chekhtman}, A., {Cheung}, C.~C.,
  {Chiang}, J., {Ciprini}, S., {Claus}, R., {Cohen-Tanugi}, J., {Costamante},
  L., {Cutini}, S., {D'Ammando}, F., {Dermer}, C.~D., {de Angelis}, A., {de
  Luca}, A., {de Palma}, F., {Digel}, S.~W., {do Couto e Silva}, E., {Drell},
  P.~S., {Drlica-Wagner}, A., {Dubois}, R., {Dumora}, D., {Favuzzi}, C.,
  {Fegan}, S.~J., {Ferrara}, E.~C., {Focke}, W.~B., {Fortin}, P., {Frailis},
  M., {Fukazawa}, Y., {Funk}, S., {Fusco}, P., {Gargano}, F., {Gasparrini}, D.,
  {Gehrels}, N., {Germani}, S., {Giglietto}, N., {Giordano}, F., {Giroletti},
  M., {Glanzman}, T., {Godfrey}, G., {Grenier}, I.~A., {Grondin}, M.~H.,
  {Grove}, J.~E., {Guiriec}, S., {Hadasch}, D., {Hanabata}, Y., {Harding},
  A.~K., {Hayashi}, K., {Hayashida}, M., {Hays}, E., {Horan}, D., {Itoh}, R.,
  {J{\'o}hannesson}, G., {Johnson}, A.~S., {Johnson}, T.~J., {Khangulyan}, D.,
  {Kamae}, T., {Katagiri}, H., {Kataoka}, J., {Kerr}, M., {Kn{\"o}dlseder}, J.,
  {Kuss}, M., {Lande}, J., {Latronico}, L., {Lee}, S.~H., {Lemoine-Goumard},
  M., {Longo}, F., {Loparco}, F., {Lubrano}, P., {Madejski}, G.~M., {Makeev},
  A., {Marelli}, M., {Mazziotta}, M.~N., {McEnery}, J.~E., {Michelson}, P.~F.,
  {Mitthumsiri}, W., {Mizuno}, T., {Moiseev}, A.~A., {Monte}, C., {Monzani},
  M.~E., {Morselli}, A., {Moskalenko}, I.~V., {Murgia}, S., {Nakamori}, T.,
  {Naumann-Godo}, M., {Nolan}, P.~L., {Norris}, J.~P., {Nuss}, E., {Ohsugi},
  T., {Okumura}, A., {Omodei}, N., {Ormes}, J.~F., {Ozaki}, M., {Paneque}, D.,
  {Parent}, D., {Pelassa}, V., {Pepe}, M., {Pesce-Rollins}, M., {Pierbattista},
  M., {Piron}, F., {Porter}, T.~A., {Rain{\`o}}, S., {Rando}, R., {Ray}, P.~S.,
  {Razzano}, M., {Reimer}, A., {Reimer}, O., {Reposeur}, T., {Ritz}, S.,
  {Romani}, R.~W., {Sadrozinski}, H.~F.~W., {Sanchez}, D., {Parkinson},
  P.~M.~S., {Scargle}, J.~D., {Schalk}, T.~L., {Sgr{\`o}}, C., {Siskind},
  E.~J., {Smith}, P.~D., {Spandre}, G., {Spinelli}, P., {Strickman}, M.~S.,
  {Suson}, D.~J., {Takahashi}, H., {Takahashi}, T., {Tanaka}, T., {Thayer},
  J.~B., {Thompson}, D.~J., {Tibaldo}, L., {Torres}, D.~F., {Tosti}, G.,
  {Tramacere}, A., {Troja}, E., {Uchiyama}, Y., {Vandenbroucke}, J.,
  {Vasileiou}, V., {Vianello}, G., {Vitale}, V., {Wang}, P., {Wood}, K.~S.,
  {Yang}, Z., \& {Ziegler}, M. 2011, {Gamma-Ray Flares from the Crab Nebula}.
\newblock {\em Science}, 331(6018), 739.

\bibitem[{Atoyan} and {Dermer}, 2004]{atoyan2004}
{Atoyan}, A. \& {Dermer}, C.~D. 2004, {Synchrotron versus Compton
  Interpretations for Extended X-Ray Jets}.
\newblock {\em \apj}, 613(1), 151--158.

\bibitem[{Breiding} et~al., 2017]{breiding2017}
{Breiding}, P., {Meyer}, E.~T., {Georganopoulos}, M., {Keenan}, M.~E.,
  {DeNigris}, N.~S., \& {Hewitt}, J. 2017, {Fermi Non-detections of Four X-Ray
  Jet Sources and Implications for the IC/CMB Mechanism}.
\newblock {\em \apj}, 849(2), 95.

\bibitem[{Breiding} et~al., 2023]{breiding2023}
{Breiding}, P., {Meyer}, E.~T., {Georganopoulos}, M., {Reddy}, K., {Kollmann},
  K.~E., \& {Roychowdhury}, A. 2023, {A multiwavelength study of multiple
  spectral component jets in AGN: testing the IC/CMB model for the
  large-scale-jet X-ray emission}.
\newblock {\em \mnras}, 518(3), 3222--3250.

\bibitem[{Cara} et~al., 2013]{cara2013}
{Cara}, M., {Perlman}, E.~S., {Uchiyama}, Y., {Cheung}, C.~C., {Coppi}, P.~S.,
  {Georganopoulos}, M., {Worrall}, D.~M., {Birkinshaw}, M., {Sparks}, W.~B.,
  {Marshall}, H.~L., {Stawarz}, L., {Begelman}, M.~C., {O'Dea}, C.~P., \&
  {Baum}, S.~A. 2013, {Polarimetry and the High-energy Emission Mechanisms in
  Quasar Jets: The Case of PKS 1136-135}.
\newblock {\em \apj}, 773, 186.

\bibitem[{Celotti} et~al., 2001]{celotti2001}
{Celotti}, A., {Ghisellini}, G., \& {Chiaberge}, M. 2001, {Large-scale jets in
  active galactic nuclei: multiwavelength mapping}.
\newblock {\em \mnras}, 321(1), L1--L5.

\bibitem[{Chartas} et~al., 2000]{chartas2000}
{Chartas}, G., {Worrall}, D.~M., {Birkinshaw}, M., {Cresitello-Dittmar}, M.,
  {Cui}, W., {Ghosh}, K.~K., {Harris}, D.~E., {Hooper}, E.~J., {Jauncey},
  D.~L., {Kim}, D.~W., {Lovell}, J., {Mathur}, S., {Schwartz}, D.~A., {Tingay},
  S.~J., {Virani}, S.~N., \& {Wilkes}, B.~J. 2000, {The Chandra X-Ray
  Observatory Resolves the X-Ray Morphology and Spectra of a Jet in PKS
  0637-752}.
\newblock {\em \apj}, 542(2), 655--666.

\bibitem[{Clautice} et~al., 2016]{clautice2016}
{Clautice}, D., {Perlman}, E.~S., {Georganopoulos}, M., {Lister}, M.~L.,
  {Tombesi}, F., {Cara}, M., {Marshall}, H.~L., {Hogan}, B., \& {Kazanas}, D.
  2016, {The Spectacular Radio-near-IR-X-Ray Jet of 3C 111: The X-Ray Emission
  Mechanism and Jet Kinematics}.
\newblock {\em \apj}, 826, 109.

\bibitem[{Feigelson} et~al., 1981]{feigelson1981}
{Feigelson}, E.~D., {Schreier}, E.~J., {Delvaille}, J.~P., {Giacconi}, R.,
  {Grindlay}, J.~E., \& {Lightman}, A.~P. 1981, {The X-ray structure of
  Centaurus A.}
\newblock {\em \apj}, 251, 31--51.

\bibitem[{Georganopoulos} et~al., 2006]{geo2006}
{Georganopoulos}, M., {Perlman}, E.~S., {Kazanas}, D., \& {McEnery}, J. 2006,
  {Quasar X-Ray Jets: Gamma-Ray Diagnostics of the Synchrotron and Inverse
  Compton Hypotheses: The Case of 3C 273}.
\newblock {\em \apjl}, 653(1), L5--L8.

\bibitem[{Giannios} et~al., 2009]{giannios2009}
{Giannios}, D., {Uzdensky}, D.~A., \& {Begelman}, M.~C. 2009, {Fast TeV
  variability in blazars: jets in a jet}.
\newblock {\em \mnras}, 395(1), L29--L33.

\bibitem[{Hardcastle} and {Croston}, 2005]{hardcastle2005}
{Hardcastle}, M.~J. \& {Croston}, J.~H. 2005, {The Chandra view of extended
  X-ray emission from Pictor A}.
\newblock {\em \mnras}, 363(2), 649--660.

\bibitem[{Hardcastle} et~al., 2007]{hardcastle2007}
{Hardcastle}, M.~J., {Kraft}, R.~P., {Sivakoff}, G.~R., {Goodger}, J.~L.,
  {Croston}, J.~H., {Jord{\'a}n}, A., {Evans}, D.~A., {Worrall}, D.~M.,
  {Birkinshaw}, M., {Raychaudhury}, S., {Brassington}, N.~J., {Forman}, W.~R.,
  {Harris}, W.~E., {Jones}, C., {Juett}, A.~M., {Murray}, S.~S., {Nulsen},
  P.~E.~J., {Sarazin}, C.~L., \& {Woodley}, K.~A. 2007, {New Results on
  Particle Acceleration in the Centaurus A Jet and Counterjet from a Deep
  Chandra Observation}.
\newblock {\em \apjl}, 670(2), L81--L84.

\bibitem[{Hardcastle} et~al., 2016]{hardcastle2016}
{Hardcastle}, M.~J., {Lenc}, E., {Birkinshaw}, M., {Croston}, J.~H., {Goodger},
  J.~L., {Marshall}, H.~L., {Perlman}, E.~S., {Siemiginowska}, A., {Stawarz},
  {\L}., \& {Worrall}, D.~M. 2016, {Deep Chandra observations of Pictor A}.
\newblock {\em \mnras}, 455(4), 3526--3545.

\bibitem[{Hardcastle} et~al., 2010]{hardcastle2010_3c171}
{Hardcastle}, M.~J., {Massaro}, F., \& {Harris}, D.~E. 2010, {X-ray emission
  from the extended emission-line region of the powerful radio galaxy 3C171}.
\newblock {\em \mnras}, 401(4), 2697--2705.

\bibitem[{Hardcastle} et~al., 2012]{hardcastle2012_3c305}
{Hardcastle}, M.~J., {Massaro}, F., {Harris}, D.~E., {Baum}, S.~A., {Bianchi},
  S., {Chiaberge}, M., {Morganti}, R., {O'Dea}, C.~P., \& {Siemiginowska}, A.
  2012, {The nature of the jet-driven outflow in the radio galaxy 3C 305}.
\newblock {\em \mnras}, 424(3), 1774--1789.

\bibitem[{Harris}, 2003]{harris2003_sed}
{Harris}, D.~E. 2003, {X-ray variability and emission process of the radio jet
  in M87}.
\newblock {\em \nar}, 47(6-7), 617--620.

\bibitem[{Harris} et~al., 2003]{harris2003}
{Harris}, D.~E., {Biretta}, J.~A., {Junor}, W., {Perlman}, E.~S., {Sparks},
  W.~B., \& {Wilson}, A.~S. 2003, {Flaring X-Ray Emission from HST-1, a Knot in
  the M87 Jet}.
\newblock {\em \apjl}, 586(1), L41--L44.

\bibitem[{Harris} and {Krawczynski}, 2006]{harris2006}
{Harris}, D.~E. \& {Krawczynski}, H. 2006, {X-Ray Emission from Extragalactic
  Jets}.
\newblock {\em \araa}, 44(1), 463--506.

\bibitem[{Kataoka} et~al., 2008]{kataoka2008}
{Kataoka}, J., {Stawarz}, {\L}., {Harris}, D.~E., {Siemiginowska}, A.,
  {Ostrowski}, M., {Swain}, M.~R., {Hardcastle}, M.~J., {Goodger}, J.~L.,
  {Iwasawa}, K., \& {Edwards}, P.~G. 2008, {Chandra Reveals Twin X-Ray Jets in
  the Powerful FR II Radio Galaxy 3C 353}.
\newblock {\em \apj}, 685(2), 839--857.

\bibitem[{Keenan} et~al., 2021]{keenan2021}
{Keenan}, M., {Meyer}, E.~T., {Georganopoulos}, M., {Reddy}, K., \& {French},
  O.~J. 2021, {The relativistic jet dichotomy and the end of the blazar
  sequence}.
\newblock {\em \mnras}, 505(4), 4726--4745.

\bibitem[{Kormendy} and {Ho}, 2013]{kormendy2013}
{Kormendy}, J. \& {Ho}, L.~C. 2013, {Coevolution (Or Not) of Supermassive Black
  Holes and Host Galaxies}.
\newblock {\em \araa}, 51(1), 511--653.

\bibitem[{Kraft} et~al., 2002]{kraft2002}
{Kraft}, R.~P., {Forman}, W.~R., {Jones}, C., {Murray}, S.~S., {Hardcastle},
  M.~J., \& {Worrall}, D.~M. 2002, {Chandra Observations of the X-Ray Jet in
  Centaurus A}.
\newblock {\em \apj}, 569(1), 54--71.

\bibitem[{Marscher} and {Jorstad}, 2010]{marscher2010}
{Marscher}, A.~P. \& {Jorstad}, S.~G. 2010, {Rapid Variability of Gamma-ray
  Emission from Sites near the 43 GHz Cores of Blazar Jets}.
\newblock {\em arXiv e-prints},, arXiv:1005.5551.

\bibitem[{Marshall} et~al., 2018]{marshall2018}
{Marshall}, H.~L., {Gelbord}, J.~M., {Worrall}, D.~M., {Birkinshaw}, M.,
  {Schwartz}, D.~A., {Jauncey}, D.~L., {Griffiths}, G., {Murphy}, D.~W.,
  {Lovell}, J.~E.~J., {Perlman}, E.~S., \& {Godfrey}, L. 2018, {An X-Ray
  Imaging Survey of Quasar Jets: The Complete Survey}.
\newblock {\em \apj}, 856(1), 66.

\bibitem[{Marshall} et~al., 2010]{marshall2010}
{Marshall}, H.~L., {Hardcastle}, M.~J., {Birkinshaw}, M., {Croston}, J.,
  {Evans}, D., {Landt}, H., {Lenc}, E., {Massaro}, F., {Perlman}, E.~S.,
  {Schwartz}, D.~A., {Siemiginowska}, A., {Stawarz}, {\L}., {Urry}, C.~M., \&
  {Worrall}, D.~M. 2010, {A Flare in the Jet of Pictor A}.
\newblock {\em \apjl}, 714(2), L213--L216.

\bibitem[{Meyer} et~al., 2017]{meyer2017}
{Meyer}, E.~T., {Breiding}, P., {Georganopoulos}, M., {Oteo}, I., {Zwaan},
  M.~A., {Laing}, R., {Godfrey}, L., \& {Ivison}, R.~J. 2017, {New ALMA and
  Fermi/LAT Observations of the Large-scale Jet of PKS 0637-752
  Strengthen the Case Against the IC/CMB Model}.
\newblock {\em \apjl}, 835, L35.

\bibitem[{Meyer} and {Georganopoulos}, 2014]{meyer2014}
{Meyer}, E.~T. \& {Georganopoulos}, M. 2014, {Fermi Rules Out the Inverse
  Compton/CMB Model for the Large-scale Jet X-Ray Emission of 3C 273}.
\newblock {\em \apjl}, 780, L27.

\bibitem[{Meyer} et~al., 2015]{meyer2015}
{Meyer}, E.~T., {Georganopoulos}, M., {Sparks}, W.~B., {Godfrey}, L., {Lovell},
  J.~E.~J., \& {Perlman}, E. 2015, {Ruling out IC/CMB X-rays in PKS 0637-752
  and the Implications for TeV Emission from Large-scale Quasar Jets}.
\newblock {\em \apj}, 805, 154.

\bibitem[{Meyer} et~al., 2018]{meyer2018}
{Meyer}, E.~T., {Petropoulou}, M., {Georganopoulos}, M., {Chiaberge}, M.,
  {Breiding}, P., \& {Sparks}, W.~B. 2018, {Detection of an Optical/UV
  Jet/Counterjet and Multiple Spectral Components in M84}.
\newblock {\em \apj}, 860(1), 9.

\bibitem[{Migliori} et~al., 2022]{mig2022}
{Migliori}, G., {Siemiginowska}, A., {Cheung}, C.~C., {Celotti}, A.,
  {Giroletti}, M., {Giovannini}, G., {Paggi}, A., \& {Liuzzo}, E. 2022,
  {Discovery of a bright extended X-ray jet in RGB J1512+020A}.
\newblock {\em \mnras}, 512(3), 4639--4659.

\bibitem[{Mingo} et~al., 2019]{mingo2019}
{Mingo}, B., {Croston}, J.~H., {Hardcastle}, M.~J., {Best}, P.~N., {Duncan},
  K.~J., {Morganti}, R., {Rottgering}, H.~J.~A., {Sabater}, J., {Shimwell},
  T.~W., {Williams}, W.~L., {Brienza}, M., {Gurkan}, G., {Mahatma}, V.~H.,
  {Morabito}, L.~K., {Prandoni}, I., {Bondi}, M., {Ineson}, J., \& {Mooney}, S.
  2019, {Revisiting the Fanaroff-Riley dichotomy and radio-galaxy morphology
  with the LOFAR Two-Metre Sky Survey (LoTSS)}.
\newblock {\em \mnras}, 488(2), 2701--2721.

\bibitem[{Omodei} et~al., 2013]{omodei2013}
{Omodei}, N., {Petrosian}, V., {Pesce-Rollins}, M., \& {the Fermi-LAT
  Collaboration} 2013, {Fermi-LAT Observation of Impulsive Solar Flares}.
\newblock {\em arXiv e-prints},, arXiv:1304.0798.

\bibitem[{Petropoulou} et~al., 2017]{petropoulou2017}
{Petropoulou}, M., {Vasilopoulos}, G., \& {Giannios}, D. 2017, {The TeV
  emission of Ap Librae: a hadronic interpretation and prospects for CTA}.
\newblock {\em \mnras}, 464(2), 2213--2222.

\bibitem[{Reddy} et~al., 2021]{karthik+21}
{Reddy}, K., {Georganopoulos}, M., \& {Meyer}, E.~T. 2021, {X-Ray-to-radio
  Offset Inference from Low-count X-Ray Jets}.
\newblock {\em \apjs}, 253(2), 37.

\bibitem[{Reddy} et~al., 2022]{karthik+22}
{Reddy}, K., {Georganopoulos}, M., \& {Meyer}, E.~T. 2022, {ATLAS-X}.
\newblock {\em \apjs},.

\bibitem[{Schwartz} et~al., 2000]{schwartz2000}
{Schwartz}, D.~A., {Marshall}, H.~L., {Lovell}, J.~E.~J., {Piner}, B.~G.,
  {Tingay}, S.~J., {Birkinshaw}, M., {Chartas}, G., {Elvis}, M., {Feigelson},
  E.~D., {Ghosh}, K.~K., {Harris}, D.~E., {Hirabayashi}, H., {Hooper}, E.~J.,
  {Jauncey}, D.~L., {Lanzetta}, K.~M., {Mathur}, S., {Preston}, R.~A.,
  {Tucker}, W.~H., {Virani}, S., {Wilkes}, B., \& {Worrall}, D.~M. 2000,
  {Chandra Discovery of a 100 kiloparsec X-Ray Jet in PKS 0637-752}.
\newblock {\em \apjl}, 540(2), 69--72.

\bibitem[{Simionescu} et~al., 2016]{sim2016}
{Simionescu}, A., {Stawarz}, {\L}., {Ichinohe}, Y., {Cheung}, C.~C., {Jamrozy},
  M., {Siemiginowska}, A., {Hagino}, K., {Gandhi}, P., \& {Werner}, N. 2016,
  {Serendipitous Discovery of an Extended X-Ray Jet without a Radio Counterpart
  in a High-redshift Quasar}.
\newblock {\em \apjl}, 816(1), L15.

\bibitem[{Tavecchio} et~al., 2000]{tavecchio2000}
{Tavecchio}, F., {Maraschi}, L., {Sambruna}, R.~M., \& {Urry}, C.~M. 2000, {The
  X-Ray Jet of PKS 0637-752: Inverse Compton Radiation from the Cosmic
  Microwave Background?}
\newblock {\em \apjl}, 544(1), L23--L26.

\bibitem[{Thimmappa} et~al., 2020]{thimmappa2020}
{Thimmappa}, R., {Stawarz}, {\L}., {Marchenko}, V., {Balasubramaniam}, K.,
  {Cheung}, C.~C., \& {Siemiginowska}, A. 2020, {Chandra Imaging of the Western
  Hotspot in the Radio Galaxy Pictor A: Image Deconvolution and Variability
  Analysis}.
\newblock {\em \apj}, 903(2), 109.

\bibitem[{Wilson} et~al., 2000]{wilson2000}
{Wilson}, A.~S., {Young}, A.~J., \& {Shopbell}, P.~L. 2000, {Chandra
  Observations of Cygnus A: Magnetic Field Strengths in the Hot Spots of a
  Radio Galaxy}.
\newblock {\em \apjl}, 544(1), L27--L30.

\bibitem[{Worrall}, 2009]{worrall2009}
{Worrall}, D.~M. 2009, {The X-ray jets of active galaxies}.
\newblock {\em \aapr}, 17, 1--46.

\end{thebibliography}

\end{document}